\documentclass[final,5p,times,twocolumn]{elsarticle} 

\usepackage{lineno,hyperref}
\modulolinenumbers[5]

\journal{Journal of \LaTeX\ Templates}

\begin{document}

\begin{frontmatter}

\title{Calibration of the ICARUS cryogenic photo-detection system at FNAL}


\author[ist1]{M.~Bonesini\corref{cor}}
\ead{maurizio.bonesini@unimib.it}
\author[ist1]{R.~Benocci}
\author[ist1]{R.~Bertoni}
\author[ist2]{A.~Chatterjee}
\author[ist3]{M.~Diwan}
\author[ist4,ist5]{A.~Menegolli}
\author[ist5]{G.~Raselli}
\author[ist5]{M.~Rossella}
\author[ist3]{A.~Scarpelli}
\author[ist2]{N.~Suarez}
\author{on behalf of the Icarus Collaboration}
\cortext[cor]{Corresponding author}

\address[ist1]{University and Sezione INFN Milano Bicocca, Milano, Italy}
\address[ist2]{University of Pittsburg, PA, USA}
\address[ist3]{Brookhaven National Laboratory, NY, USA}
\address[ist4]{University of Pavia, Pavia, Italy}
\address[ist5]{Sezione INFN Pavia, Pavia, Italy}




\begin{abstract}
The calibration of the ICARUS photo-detection system 
is based on a low power laser diode at 405 nm. 
Laser pulses arrive to one optical switch and then are sent to 36 UHV flanges, 
by 20 meters long optical patches. Light is then delivered to the ten PMTs
connected to a single flange, by 7m long injection optical patches. Extensive tests 
of the used components and care in the design of the optical system have guaranteed to each PMT
a sizeable calibration signal with minimal distortion, with respect to the original one.
Gain equalization of PMTs has reached a 1 $\%$ resolution. In this procedure 
data from background photons were also used.
The distribution of the PMTs' signal arrival time has a distribution with a 
resolution less than
1 ns, thus allowing a good determination of the absolute event 
timing.  
The status of the laser calibration 
system with its possible upgrades will be reported. 

\end{abstract}


\end{frontmatter}


\section{Introduction}
ICARUS T600  \cite{T600} is presently 
used as far detector of the Short Baseline Neutrino (SBN) program at 
Fermilab (USA) to search for a possible LSND-like sterile neutrino
signal at $\Delta m^{2} \sim $ O(eV$^{2}$ ) \cite{Antonello:2015}. 
It is made of two identical cryostats, filled with about 760 tons 
of ultra-pure liquid Argon. Each cryostat houses two TPCs with 1.5 m 
maximum drift path, sharing a common central cathode.
Charged particles interacting in liquid argon produce both scintillation 
light at 128 nm and ionization electrons.
ICARUS  is placed at shallow depth 
on the Booster Neutrino beam (BNB). To reduce the cosmic ray background, in 
addition to a full coverage cosmic ray tagger (CRT), a system based on 360 
large area Hamamatsu R5912-MOD photomultipliers (PMTs) \cite{Raselli:2020}, directly immersed 
in liquid Argon, is used to detect  scintillation 
light. 
To exploit the BNB  bunch structure, for futher rejection of out-of-bunch cosmic
events, an overall time resolution $\sim 1$ ns is  needed. This  and the
trigger system
require an accurate calibration in gain and time 
of each PMT. 
\vskip -3.5cm
\section{The laser calibration system}
\begin{figure*}
\centering
\vskip -1cm
\includegraphics[width=\linewidth]{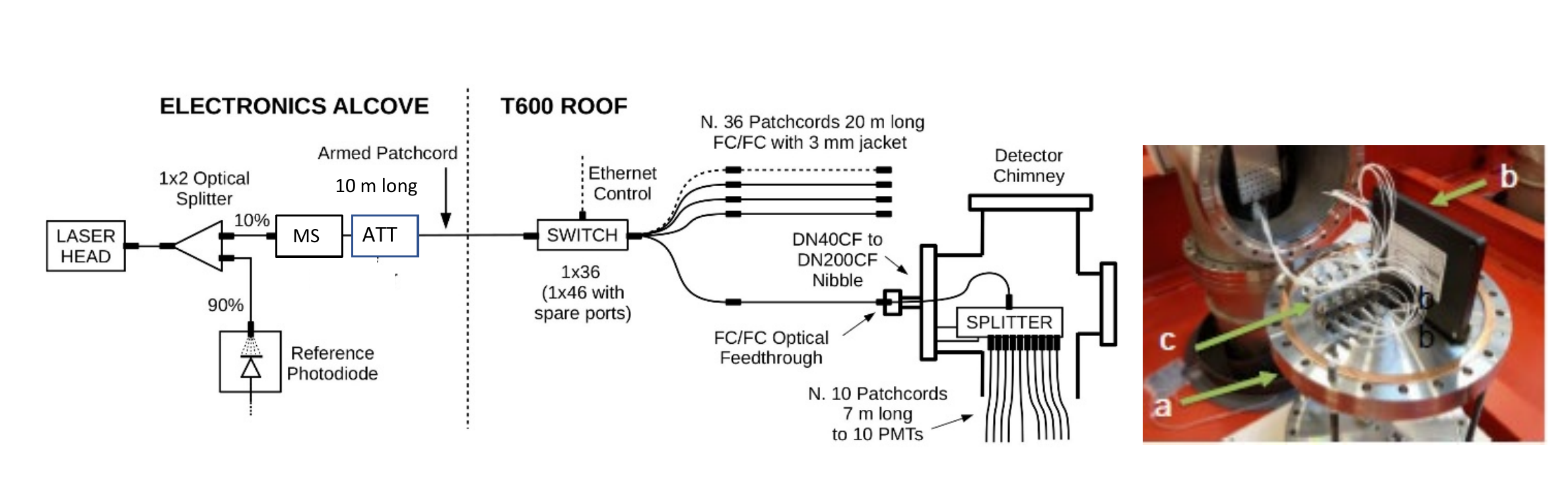}
\vskip -0.7cm
\caption{Left panel: schematic layout of the laser system realized for  
calibration of the 360 PMTs of the Icarus T600 detector.
	Right panel: image of one UHV flange (a) with the 1x10 splitter (b) 
	and the FC/FC patch panel (c).
}
\label{fig:layout}
\end{figure*}

The PMTs  equalization may be performed by using fast
laser pulses, as done in previous experiments such as Borexino at LNGS 
\cite{borex} and HARP at CERN PS \cite{harp}. The light pulse from a PLP10 Hamamatsu laser diode 
is sent to each PMT via a
distribution system  including an Arden
Photonics  Mode Scrambler (MS), a DD-100 OZ/Optics 
programmable attenuator(ATT), a 10 m  armed fiber
patch cable, followed by one 1x46 Agiltron optical switch (where  10
channels are spares). From this 
20 m fiber patch cords go to  VACOM UHV optical feedthroughs
(36) on CF40 flanges, mounted on CF40-CF20 nibbles, as shown in figure 
\ref{fig:layout} . Inside the T600 tank, 1x10 Lightel fused fiber splitters 
attached to each nibble deliver the input laser signal to the window of 
each PMT.  
While an optical switch delivers the laser signal to a single output
channel, a 1xN optical switch distributes the input signal among 
N output channels. 
The light delivery system must have a minimal spread in channel-to-channel 
total delay
($\Delta t$) and delivered signal power in front of each PMT.  
In addition laser pulses have to 
be delivered to the PMT's photocathode with a minimal attenuation and
without a deterioration of the original
timing characteristics.
\vskip -3.5cm
\section{Optical components characterization}
To keep costs low, commercial optical components (fibers, optical switches,
fused optical splitters, ...) were used.
As these components are easily available only 
at Telecom wavelengths ($\sim 850 \div 1300$ nm), we had to characterize 
their behaviour at 400 nm  with a dedicated test bench at INFN Milano Bicocca. 
To minimize alignment problems multimode fibers (MM) instead of single
mode ones (SM) were used. 
At this point, the remaining main task 
was to found  a suitable optical switch and fused  1x10 optical splitters
with channel-to-channel minimal spread in insertion losses (IL) and delay.
The choice was to use an Agiltron  1x46 optical switch and 
Lightel 1x10 fused optical splitters. 
In the optical switch, the FWHM of the laser pulses is increased at most by 
$3 \%$ and the insertion loss (IL) is less than 0.5 dB, with a cross talk
$\sim 50$ dB. Channel-to-channel output uniformity is better than $ 10 \%$ 
and the RMS of internal time delays is $\sim$ 10 ps. 
Instead, a typical  splitter increases laser pulse FWHM of less than 
$4 \%$ and has an IL less than 3 dB. For each splitter, the signal uniformity in the 10
output channels  is less than $5 \%$ aside one leading channel, by construction, that has 
to be attenuated.
Average internal delays of the 36 splitters are within 80 ps (aside two, 
belonging to a different batch).

UHV optical feedthroughs   convey the
laser pulse inside the Icarus cryogenic tank. 
The adopted solution  from VACOM, with a MM
50 $\mu$m core fiber,  had a
measured transmission  around 80 \%, introduced an additional
delay $\sim 100 $ ps and a negligible  additional time dispersion (FWHM).
To adapt these
flanges to the existing T600 chimneys CF40-CF200 nibbles are used,
see figure \ref{fig:layout} for details.
As  the  splitters have to work, inside the T600 chimneys,
at a slightly lower temperature than the T600 roof 
and the 7m injection
patches at cryogenic temperatures, inside the LAr bath, tests were done
to assess their dependence from temperature, using a
Lauda thermal machine (precision $0.1$ $^{0}$C) in the range $ 0 \div 50 \ 
^{0}$C  and  with $LN_2$ bath inside a dewar, see reference \cite{Bonesini:2020}. 

\vskip -3.5cm
\section{System performances}
At the PMTs' front face, a total delay of the laser signal ($\Delta t \sim 250$ ns) 
with a channel-to-channel spread estimated
at less than 200 ps was measured both in situ and in laboratory. 
Up to the UHV flanges a $4.59 \pm 0.16$ dB optical
signal attenuation
was measured. The attenuation of the 7 m injection patches was measured as
$0.61 \pm 0.16 $ dB for the full 410 sample (best 360 used).
\begin{figure}
\centering
\includegraphics[width=\linewidth]{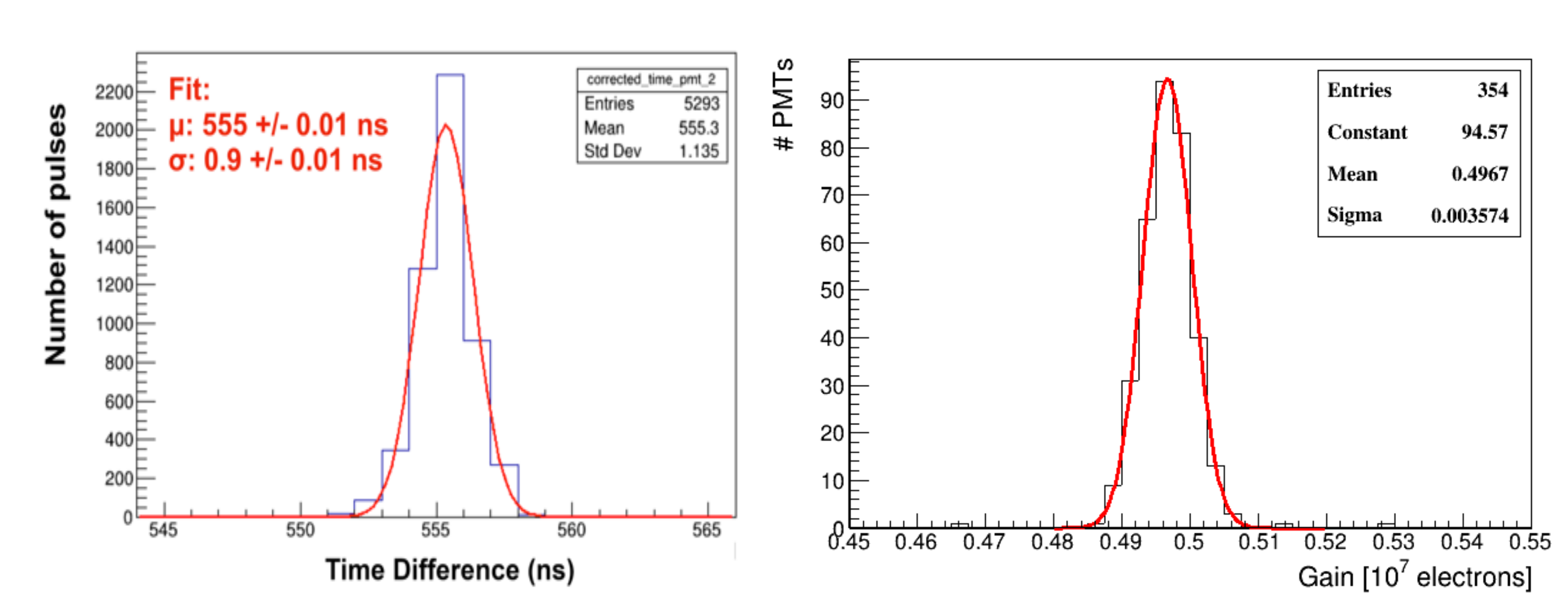}
\caption{Left panel: time difference $\Delta T$ distribution. Right panel: PMTs gain distribution after equalization. }
\label{fig:equa}
\end{figure}
For calibration purposes, charge spectra from PMTs were 
recorded integrating laser signals over a 100 ns
window. For each PMT data were taken at different voltages to allow the 
determination of the gain vs voltage curve and thus provide the voltage needed 
for a nominal gain set at $5 \times 10^6$. A fine tuning with background 
single photoelectrons is then applied to reduce the relative gain spread to 
less than $1 \%$, as shown in figure \ref{fig:equa}. 
In the same figure the  distribution of the time difference $\Delta T$ 
between the trigger and the PMT signal is shown. Timing of PMTs signals 
is given by $T_{PMT} = T_{TRIG} + \Delta T$, where the trigger pulse is 
synchronized with the beam RF. 
Laser pulses permit the monitoring of possible variation of $\Delta T$ along
the data taking and between different channels. With the obtained 
resolution of 1 ns,
it is possible to determine the absolute timing of collected events, for 
trigger and cosmic rays rejection purposes.
   
As the PMTs' response to calibration pulses is sizeable, as measured in situ
using the optical attenuator, the replacement of the optical switch (10 PMTs 
calibrated in a single run) with a custom 1x36 optical splitter (360 PMTs 
calibrated in the same run) is foreseen in the next future. This will allow
a much faster laser calibration procedure. 
\vskip -3.5 cm
\section{Conclusions}
The light detection system of ICARUS T600 is fully operational allowing a 
regular data taking. Calibration of the PMTs' gain has a spread less then $1 \%$
while PMTs timing has a resolution better than 1 ns.
Some upgrades are foreseen to improve
the automation and reduce the  time needed for the laser calibration procedures.
\vskip -3.5 cm
\section*{Acknowledgements}
This work was supported by EU Horizon 2020 grant Agreements n. 74303, 822185, 858189 and 101003460.

\bibliography{paper_v2}

\end{document}